# Amplified Emission by Atoms and Lasing in Photonic Time Crystals

Mark Lyubarov[1,2,3], Yaakov Lumer[1,2], Alex Dikopoltsev[1,2], Eran Lustig[1,2],

Yonatan Sharabi[1,2], Mordechai Segev[1,2,4]

[1] Physics Department, Technion – Israel Institute of Technology, Haifa 32000, Israel

[2] Solid State Institute, Technion – Israel Institute of Technology, Haifa 32000, Israel

[3] Physics and Engineering Department, ITMO University, St. Petersburg 197101, Russia

[4] Department of Electrical and Computer Engineering, Technion – Israel Institute of Technology, Haifa 32000, Israel

## Abstract

Photonic Time Crystals (PTCs) – dielectric media with their refractive index modulated periodically in time – offer new opportunities in photonics arising from time-reflections and momentum bandgaps. Here, we study the emission of light from a radiation source inside a PTC. We solve the general classical and quantum mechanical models of emission in a temporally-varying medium, and find that radiation is always exponentially amplified when associated with the momentum gap, whether initiated by a macroscopic source, an atom, or by vacuum fluctuations, drawing the amplification energy from the modulation. The radiation linewidth becomes narrower as time advances, and is centered in the middle of the momentum gap. We calculate the spontaneous decay rate of an atom embedded in a PTC and show that it vanishes at the band edge due to low density of photonic states. Finally, we propose the concept of non-resonant tunable PTC lasers.

Photonic Time Crystals (PTCs) are dielectric media whose refractive index experiences large ultrafast periodic variations in time [1-5]. Generally, a wave propagating in a medium undergoing an abrupt change in the refractive index experiences time-reflection and time-refraction. In a spatially-homogeneous time-varying medium, momentum is conserved whereas energy is not, hence the frequencies of the time-reflected and time-refracted waves change due to the index variation. The time-reflection is especially interesting, because causality imposes that wave reflected from the temporal interface propagates backward in space (rather than in time, which is still unfortunately impossible) [6]. Time-refraction and reflection, fascinating wave phenomena on their own right, occur in many wave systems in electromagnetic (EM) waves [7], water waves [8], acoustic waves [9], elastic waves [10], and in principle can occur in any wave system. Periodic modulation of the refractive index makes these time-reflections interfere, giving rise to a band structure with bands and gaps in the momentum [11,12]. The most important feature of PTCs is the existence of a bandgap in momentum, because the modes associated with this momentum gap have two solutions, where the mode amplitude grows or decays exponentially, and both solutions are physical. The exponentially-growing gap mode offers an avenue for amplification of radiation, drawing energy from the modulation. The exponential growth of the gap modes is non-resonant: it occurs for all wavevectors associated with the momentum gaps.

The dispersion relation of PTCs, exhibiting bands and gaps, makes PTCs analogous to spatial photonic crystals (SPCs), where the refractive index is periodically modulated in space. However, despite the similarity, there are fundamental differences between SPCs and PTCs: SPCs are stationary in time, hence energy conservation governs most processes, whereas in PTCs energy is not conserved, and causality dictates the dynamics in the system. On the other hand, waves propagating in SPCs exchange momentum with the spatial lattice, whereas in spatially-homogeneous PTCs, momentum is conserved.

Apart from the temporal reflection and refraction, and the momentum band structure, the abrupt temporal modulation of the EM properties also opens up a number of new possibilities such as a frequency conversion [4,5,12], photon pair creation from vacuum [13-17], topological temporal edge states [18], antireflection temporal coatings [19], extreme energy transformations [20], and localization combined with amplification in temporally-disordered media [21]. Experimentally, time-refraction was already observed in photonics [4], while time-reflection was thus far observed only with water waves [8] and elastic waves [10]. This is due to the highly demanding requirements for detectable time-reflections: the change in the refractive index should act as a "wall", in analogy to a spatial interface in SPC causing Fresnel reflection. Thus, the refractive index should vary on the order of unity at a rate comparable to the frequency of the EM wave experiencing the index modulation. For light in the near infrared, the modulation should be at a few femtosecond rates with an index change of >0.1, which is extremely difficult to realize in experiment. However, the recent progress with epsilon-near-zero materials exhibiting ultrafast carrier dynamics [4, 22-25] has brought these ideas close to meeting such requirements, to the extent that experimental observation is anticipated in the near future [26].

The existence of momentum bands and gaps in a PTC raise fundamental questions on the emission of light by a radiation source embedded in a PTC. Thirty-five years ago, the analogous study in spatial photonic crystals led to the discovery of the inhibition of spontaneous emission in the EM bandgap of SPCs [27]. Since, generally in lasers, spontaneous emission into non-lasing modes is always considered as inevitable loss, this discovery had enormous significance, and, in fact, gave birth to the field of Photonic Crystals. One of its major consequences made it possible to realize threshold-less lasing into a defect mode in the photonic bandgap of SPCs [28, 29]. In this spirit, it would be interesting to explore the radiation emitted by a radiation source embedded in a PTC. Naturally, this can have many fundamental implications and offer exciting applications.

Here, we formulate the quantum mechanical theory describing the emission of light by atoms in excited state and the classical theory of radiating dipoles embedded in PTCs. We show that radiation is always exponentially amplified when associated with the momentum gap, and its linewidth is becoming narrower with time. This discovery allows proposing the concept of non-resonant tunable PTC lasers, drawing their energy from the modulation.

Consider a source of radiation embedded in a PTC – a homogeneous lossless dielectric material with permittivity $\varepsilon(t)$ modulated externally in a determined way with a period $T$, as sketched in Fig. 1a. First, we assume that the radiation source is a point dipole, i.e., acts as a temporally-oscillating point current $\boldsymbol{j} = -i\omega\delta(\boldsymbol{r})\boldsymbol{d}_0 e^{-i\omega_0 t}$. The emission pattern of a point dipole in SPC is well known: if its frequency $\omega_0$ resides in an "allowed" frequency band, it excites the respective Bloch modes in the bulk of the photonic crystal, whereas if $\omega_0$ resides within a photonic bandgap - no radiation occurs. But how does such a dipole emission behave in a PTC, where the bands and the bandgaps are defined with respect to the wavevector, rather than the frequency? As we show in Fig. 1b, the emission of a point dipole embedded in a PTC always grows exponentially with time, drawing its energy from the modulation. This is because the medium is homogeneous, hence a point dipole always radiates into all modes, including those associated with the momentum gap, where the Floquet eigenmodes have amplitudes that grow exponentially with time, as will be explained below. This growth barely depends on the frequency of the dipole, but it strongly depends on the amplitude of the permittivity modulation. Namely, the larger the modulation, the sooner the growth takes place and the steeper it is. This can be explained by the fact that larger modulation amplitude of the permittivity allows to import more energy in the system within a modulation cycle, since energy can be imported only when the permittivity varies. Another way to describe this is the band-structure of the PTC, Fig. 1c.

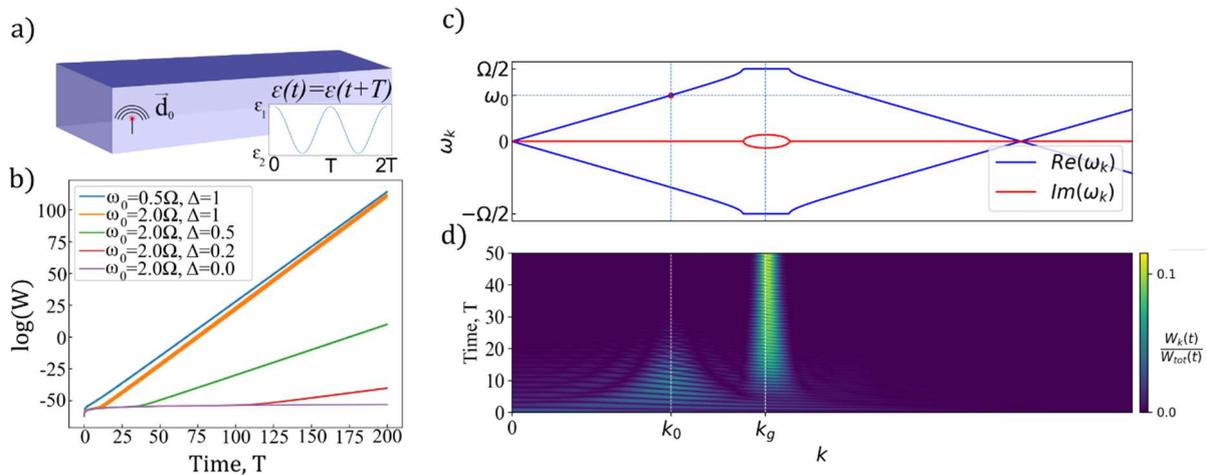

Figure 1. Emission by a point dipole embedded in a PTC. a) A sketch of the PTC, with permittivity varying according to $\varepsilon(t) = \varepsilon_{av} + \frac{\Delta}{2}\cos(\Omega t)$, $\Omega = 2\pi/T$, with a dipole antenna inside. b) Exponential growth of EM energy associated with the dipole emission for different dipole frequencies $\omega_0$ and modulation amplitudes. c) Complex dispersion relation (bandstructure) of the PTC for $\varepsilon(t) = 2 \pm 0.5\cos(\Omega t)$. d) Power spectrum of dipole emission vs wavenumber, as it evolves with time. In each moment of time the spectrum is normalized by the total radiation power. $k_0$ is a wavenumber of the mode resonantly excited by the dipole with frequency $\omega_0$: $\omega_k(k_0) = \omega_0$. The emission linewidth initially occupies the whole bandgap, but becomes narrower with time reflecting the stronger emission at middle of the gap. The horizontal axes in c), d) coincide.

Figure 1c shows the dispersion relation of the PTC, $\omega_k(k)$, where $\omega_k$ is a Floquet frequency. As shown there, in certain ranges of wavenumber $k$, $\omega_k$ has a nonzero imaginary part. This imaginary part is responsible for gain and loss in the system during the modulation cycle, and we refer to these ranges of $k$ as to the PTC bandgaps. Generally, the larger the modulation of $\varepsilon$, the wider PTC bandgaps it opens, and the larger $|Im(\omega_k)|$ in these bandgaps. The power spectrum (in $k$) of the dipole emission and its evolution with time are depicted in Fig. 1d. Initially, the point dipole with frequency $\omega_0$ efficiently excites all the eigenmodes with proper wavenumber $k_0 = \omega_0\langle n\rangle/c$, where $\langle n\rangle$ is the effective refractive index – some mean value through the modulation cycle, and $c$ is the speed of light in vacuum. This is because these waves lie on dispersion curve, and thus are perfectly phase-matched. But, within a few oscillation cycles, the gap modes start to dominate in the spectrum, even if $k_0$ does not belong to the gap. These modes are not phase-matched with the dipole frequency, but they nevertheless grow exponentially in time (extracting energy from the modulation), which overshadows any phase-matching. ***In the next section we explain in more details the possibility of exponentially growing emission by rigorously solving the Maxwell equations and finding the eigenmodes of the PTC analytically.***

First, we consider an empty PTC medium (no radiation source), and derive the eigenmodes, and then add an arbitrary radiation source. Starting with curl Maxwell equations with $\varepsilon = \varepsilon(t), \mu = 1$, we write the wave equation for the magnetic field as

$$(\partial_t(\varepsilon(t)\partial_t) + c^2 k^2)\boldsymbol{H_k} = 0 \tag{1}$$

where we use a Fourier transform in space, since the system is homogeneous and $k$ is a good quantum number. Physically, this means that the eigenmodes are shaped as plane-waves, defined by their wavenumber $k$. For each wavenumber $k$ this equation has two Floquet eigenmodes:

$$H_k^{1,2}(t) = H_{k0}(t)e^{i\omega_k^{1,2}t} \tag{2}$$

where $\omega_k^{1,2}$ are Floquet quasi-frequencies and $H_{k0}(t)$ is a periodic function in time, constructed from harmonics of the modulation period $T$. Since $\varepsilon$ is real (the medium is lossless), if $H(t)$ is an eigenmode, i.e., solution of (1), so is $H^*(t)$, which means $\omega_k^1 = -\omega_k^2 = \omega_k$. Solving for the dispersion relation, we find that the dispersion curve forms a band structure, Fig. 1c. In the bands – the frequency $\omega_k$ is real and the two modes are oscillating at the same frequency, whereas in the band gaps $\omega_k$ has an imaginary part, with one mode exponentially growing with time while the other mode is exponentially decaying. Up to this point, the properties of the system seem to be analogous to the properties of Photonic Crystals. The interesting difference here is that the exponentially growing modes in the gaps of PTCs are not unphysical (as they are in dielectric SPCs). Rather, in PTCs radiation sources can actually couple to these diverging modes, and lead to the extraction of energy from the modulation. To explore this, we add to Eq. (1) a radiation source associated with a temporally-dependent current density $j(t)$:

$$(\partial_t(\varepsilon(t)\partial_t) + c^2 k^2)\boldsymbol{H}_k(t) = 4\pi i c \boldsymbol{k} \times \boldsymbol{j}_k(t) \tag{3}$$

where $\boldsymbol{j}_k(t)$ is a Fourier $\boldsymbol{k}$ component of current. Physically, the field $\boldsymbol{H}_k(t)$ is the response of the medium to this current. We can express it in a general form through the Green function of the system as

$$\boldsymbol{H}_k(t) = 4i\pi c \int_{-\infty}^{\infty} G_k(t,t') \boldsymbol{k} \times \boldsymbol{j}_k(t') dt' \tag{4}$$

$$(\partial_t(\varepsilon(t)\partial_t) + c^2 k^2) G_k(t,t') = \delta(t-t') \tag{5}$$

and then express this Green function through the system's eigenmodes from Eq. (2)

$$G_k(t,t') = \begin{cases} 0, & t < t' \\ \dfrac{H^2{}_k(t') H^1{}_k(t) - H^1{}_k(t') H^2{}_k(t)}{\varepsilon(t')(H^2{}_k(t')\partial_{t'} H^1{}_k(t') - H^1{}_k(t')\partial_{t'} H^2{}_k(t'))}, & t > t' \end{cases} \tag{6}$$

The Green function $G_k(t,t')$ represents the response of the medium at time $t$ to a single homogeneous "flash" at time $t'$. The detailed derivation of Eq. (6) is provided in the Supplementary Information [30]. It should be noted that the analogous spatial Green function for SPCs, $G_\omega(x,x')$, is defined similar to Eq. (5). A closer look at the expression (5) reveals that - in the momentum band gap, where $Im(\omega_k) \neq 0$ - the medium responds with exponentially growing emission even to the slightest flash of radiation emitted from the current source. This seemingly counterintuitive feature is a consequence of the lack of energy conservation in the medium. In fact, the energy deposited into the exponentially growing gap modes comes not from the source but from the external modulation of the medium.

We can qualitatively describe the exponentially growing response in a PTC with the help of Fig. 2. The figure shows the difference between excited gap modes in SPC and in PTC, where the excitation in the SPC is by a point source in real-space, and the excitation in the PTC is by a "flash" in time. Mathematically, the solution of Eq. (5) has two degrees of freedom, and therefore should be expressed through two eigenmodes on either side of the excitation moment $t = t'$, and stitched with two stitching conditions at this point. To understand the physical consequences without solving the equation, we must use the physical constraints in both cases. In the case of SPC, Fig. 2a, the solution must obey energy conservation, so only evanescent waves are allowed on either side of the excitation point in space. Hence, the response to the excitation at a frequency in the gap of a photonic crystal is evanescent waves. On the other hand, in the PTC, Fig. 2b, two of the four modes are propagating back in time, therefore cannot be excited – restricted by causality. At the same time, energy conservation does not apply for PTCs, thus the Green function must be expressed with two forward-propagating waves in time, one of which is exponentially decaying and the other exponentially growing.

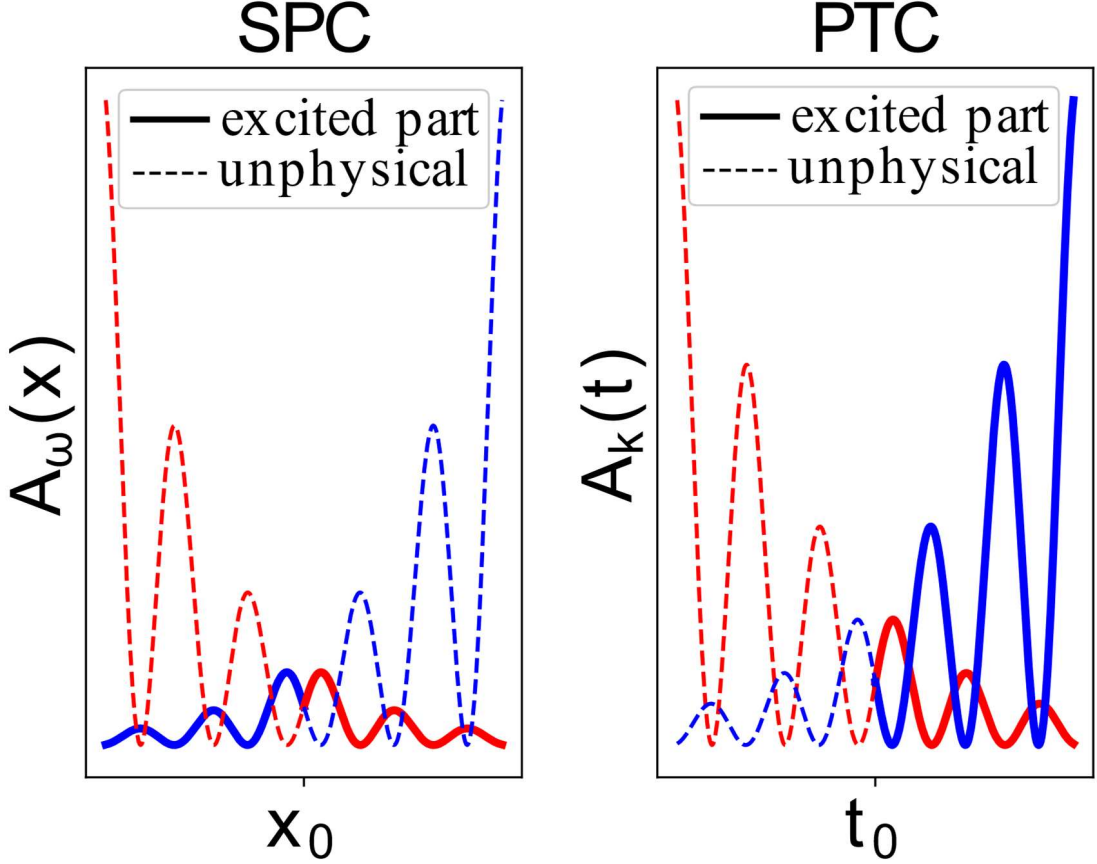

**Fig. 2. Comparison between the excitation of gap modes in SPC and PTC**. (a) The 1D SPC is excited by a point source at position $x_0$ and emits at a given frequency within the photonic bandgap. The source can couple only to the spatially- evanescent part on either side of $x_0$, due to energy conservation. (b) In the PTC, the source is a flash at $t_0$, and it can excite only to the parts of the modes that evolve forward in time, as dictated by causality. One of these two modes is exponentially growing in time.

This analysis explains the exponentially growing dipole emission in a PTC. The dipole excites the gap modes, which, once excited, grow exponentially regardless the dipole, even when mismatched. The key issue here is that we assumed a **point** dipole. Hence, because it is a point, it excites modes with all $k$, including the exponentially growing gap modes. Thus, any point source in a PTC results in exponentially growing emission, even when the excitation is a single flash in time. The emission from this flash will grow exponentially, drawing energy from the modulation. In the next section we describe the radiation in PTCs in a quantum picture. We describe the emission by an atom initially in the excited state and the influence of the PTC bandstructure on the decay time to the ground state, and discuss the evolution of the photon numbers in the modes at the PTC bandgap.

The problem of quantization of EM waves in a time-varying medium is related to the dynamical Casimir effect (DCE) [31], since a periodic change of the permittivity of a dielectric medium installed in a cavity effectively acts as a periodic change of the boundary conditions, such as the distance between the mirrors forming the cavity [14,15,31-35]. The PTC Hamiltonian we are using is

$$H_f = \hbar \sum_k \frac{ck}{n(t)} \left( \frac{\frac{n(t)}{n_r} + \frac{n_r}{n(t)}}{2} (a_k^\dagger a_k + a_{-k}^\dagger a_{-k}) + \frac{\frac{n(t)}{n_r} - \frac{n_r}{n(t)}}{2} (a_k a_{-k} + a_k^\dagger a_{-k}^\dagger) \right) \quad (7)$$

where $n(t) = \sqrt{\varepsilon(t)}$ is the time-varying refractive index, and $n_r$ is the mean value of refractive index obtained by averaging through one modulation cycle. This Hamiltonian is derived in the Supplementary Information [30] following the quantization procedure described in [36], and it qualitatively coincides with [37, 38]. It follows our intuition gained in the classical case: it is time-dependent through $n(t)$, it conserves momentum, $[H_f, \sum_k \boldsymbol{k} \, a_k^\dagger a_k] = 0$, but does not conserve the number of photons. The structure of the Hamiltonian (7) allows us to describe the dynamics of the free field for each photon pair $\{\boldsymbol{k}, -\boldsymbol{k}\}$ separately. Doing so, we find that the resulting dynamics agrees with classical case: for modes $k$ associated with the band of the PTC, the expectation value of the number of photons $\langle N_k \rangle(t) = \langle \psi(t) | a_k^\dagger a_k + a_{-k}^\dagger a_{-k} | \psi(t) \rangle$ oscillates near some constant value, whereas if $k$ belongs to the PTC bandgap, $\langle N_k \rangle$ grows exponentially with time at the same rate as in the classical case. The observation holds even if the initial state is vacuum, which is in agreement with known results on the photon generation rate in DCE under resonant conditions, $\langle N_k \rangle \propto \sinh(\gamma t)$ [34]. The periodic variation of $n(t)$ allows us to introduce the Floquet eigenmodes $|\psi_k(t)\rangle = e^{-i\omega_k t} |\phi_k(t)\rangle$ of the Hamiltonian (7), with $\omega_k$ being the Floquet eigenfrequency. In the band, $\omega_k$ coincides with the Floquet frequency calculated in the classical analysis, and the Floquet eigenstates do not differ much from the corresponding Fock states. On the other hand, in the bandgap, the eigenstates of the quantum Hamiltonian cannot exist: by correspondence with the classical case, $\langle N_k \rangle$ in the gap eigenmodes should grow exponentially, which is impossible with Hermitian Hamiltonians, such as the one in Eq. (7). The absence of eigenstates in the gap brings complexity in studying the dynamics of the excited atom due to light-atom interaction, which is described below, but the exponential growth of the number of photons in the momentum gap and the classical/semiclassical intuition allow to make safe statements on the dynamics in this unusual quantum system.

To describe the emission from excited atoms in PTC, we add the atomic and the interaction parts in Hamiltonian (7)

$$H = H_f + H_a + H_{int} \tag{8}$$

$$H_a = \hbar \omega_0 \sigma_z \tag{9}$$

$$H_{int} = \sum_k \frac{\hbar g_k}{\varepsilon(t)} (a_k + a_k^\dagger)(\sigma_+ + \sigma_-) \tag{10}$$

where we assume a two-level atom and dipole interaction. We first want to analyze what happens with an initially excited atom interacting with the vacuum field. In the stationary case (of a static medium) this problem is solvable analytically, and results in the exponential decay of the atom from the excited state to the ground state, known as spontaneous emission. In a PTC, no analytic solution is feasible, at least because vacuum is not an eigenstate of the light field. The bigger problem, however, is the fact that the number of photons in the initially empty gap modes will grow exponentially regardless of the atom. This means that the atom emission into these modes cannot be clearly divided into spontaneous and stimulated emissions: the rate of transitions will grow with time as a consequence of the photons already created by the PTC. Apart from stimulated emission, stimulated absorption also takes place. The overall dynamics of an atom interacting with gap modes does not result in the decay of the excited state to the ground state, but rather it is stabilizing at 50% probability in the excited and 50% in the ground state, with a constantly growing number of photons. As in the classical case, the growth of the number of photons barely depends on the frequency of atomic transition, and, even if the two-level atom is not in resonance with the middle of the momentum-gap, i.e. $\omega_0 \neq \Omega/2$, the emission into gap modes eventually governs the dynamics of the atom.

It is now natural to ask if there are any particular circumstances under which we can still talk about spontaneous emission (in the usual sense of being induced by quantum fluctuations with no photons around) in a PTC, and if there are, what are the physical consequences. This question can be answered partially by addressing the Floquet modes associated with the band, while the influence of the gap modes is omitted. This assumption can be justified if decay time of the atom is shorter than inverse growth rate of the number of photons in the gap modes $\tau_{sp} < 1/Im(\omega_k)$, or if the PTC with the embedded atom is placed in a resonator with all resonator eigenmodes residing inside the PTC bands (rather than in the gaps). In this case, we show in the Supplementary Information [30] that the spontaneous emission rate is

$$\gamma = \frac{V}{\hbar^2 \pi} \sum_m |V_{fi}^m|^2 \frac{k_m^2}{\left|\frac{\partial \omega_f}{\partial k}\right|_{k=k_m}} \tag{11}$$

where

$$V_{fi}^m = \frac{1}{T} \int_0^T \langle \phi_f(t)|H_{int}(t)|\phi_i(t)\rangle e^{im\Omega t} dt \tag{12}$$

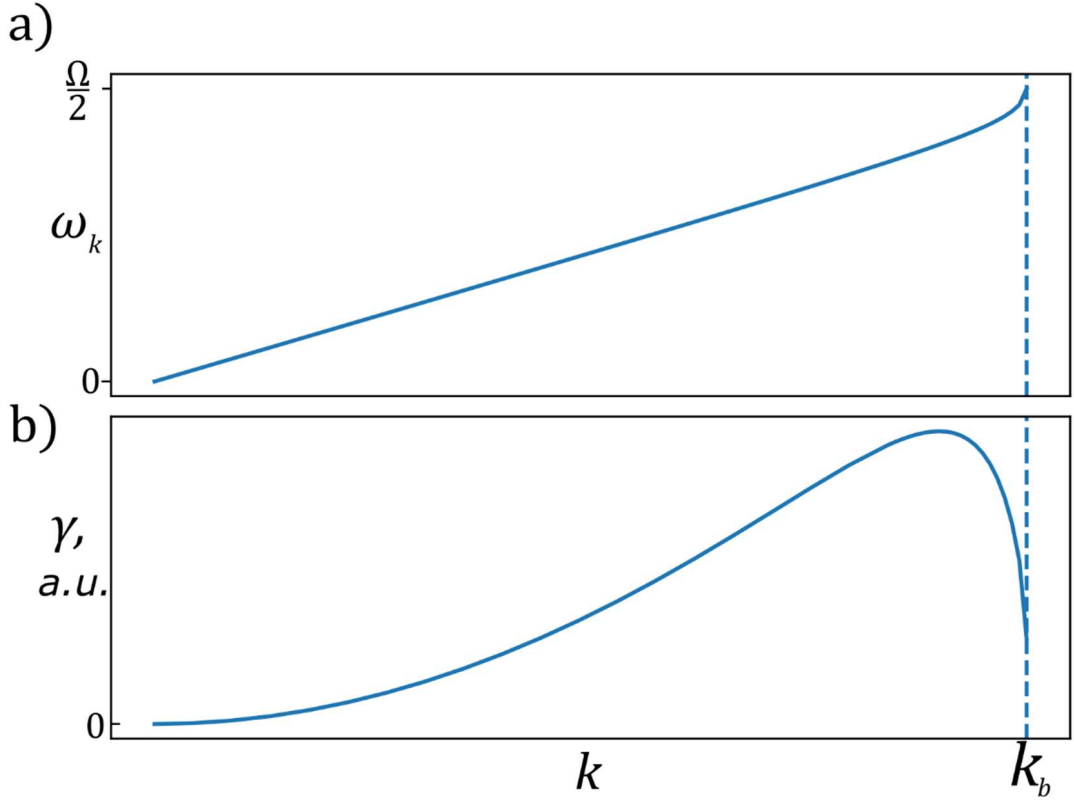

*Figure 3. Spontaneous emission rate into the Floquet modes associated with the first band. a) Dispersion in the first band of the PTC. b) Spontaneous emission rate in the same first band. Starting at low wavenumber, the emission rate increases, reaches a maximum but then declines and goes to zero at the band edge, $k_b$, where the bandstructure is curved.*

is the coupling constant between the initial and final Floquet eigenstates through $H_{int}$, and $k_m: \omega_f(k_m) = \omega_0 + m\Omega$ is the wavenumber of the mode corresponding to $m^{th}$ harmonic of the atomic transition. Analyzing the wavenumber dependence of the emission rate $\gamma(\omega)$ within the band, Eq. (11) shows that there are two competing effects when we move closer to the band edge. On the one hand, the closer to the band gap the larger the amplitude of the field oscillations in Floquet eigenstate, which is manifested in a larger $|V_{fi}^m|$. On the other hand, the density of photon states, $\rho \propto k^2 \left(\frac{\partial \omega}{\partial k}\right)^{-1}$, vanishes near the band edge, which can be seen by the vertical slope of the dispersion relation near the gap in the Fig. 1c. Surprisingly, the second effect is stronger than the first. Thus, the spontaneous emission rises up to some wavenumber but then vanishes completely near the very edge, where the curvature of the band structure determines the outcome. The implication is very intriguing: despite the fact that the Floquet modes have larger oscillations closer to the band edge, which naturally increases the strength of light-matter interaction, the emission rate at the edge goes to zero because there are no states to radiate into. Thus, an "atom" or a nano-antenna with directional emission at the band edge would stay in the excited state forever, unable to relax to the ground state through spontaneous emission.

In conclusion, we formulated the foundations of light-matter interaction in a one-dimensional photonic time-crystals. The presence of a gap in the momentum alters light-matter interactions in a profound way, bringing to question foundational issues such as the meaning of spontaneous and induced emission in such media, and the lifetime of an atom in excited states. The exponential growth of the energy in the modes associated with the PTC gap and the non-monotonous growth rate raise the exciting of idea of PTC lasers, which extract their energy from the modulation. The main idea is that a controllable periodic change of the permittivity can give rise to coherent radiation from an almost arbitrary source, and, under some conditions, the emission can be shaped into pulses by designing the permittivity modulation.

# References

1. F. R. Morgenthaler, Velocity Modulation of Electromagnetic Waves, *IEEE Trans. Microw. Theory Tech*. **6**, 167–172 (1958).

2. R. Fante, Transmission of electromagnetic waves into time-varying media. *IEEE Trans. Antennas Propag.* AP-19, 417 (1971).

3. C. L. Jiang, Wave propagation and dipole radiation in a suddenly created plasma." IEEE Trans. Antennas Propag. AP-23, 83 (1975).

4. Y. Zhou, M. Z. Alam, M. Karimi, J. Upham, O. Reshef, C. Liu, A. E. Willner, R. W. Boyd, Broadband frequency translation through time refraction in an epsilon-near-zero material. *Nat. Commun*. **11**(1), 2180 (2020).

5. J. R. Zurita-Sánchez, P. Halevi, J.C. Cervantes-Gonzalez, Reflection and transmission of a wave incident on a slab with a time-periodic dielectric function $\epsilon$ (t). *Phys. Rev. A* **79** 053821(2009).

6. J. T. Mendonça, P. K. Shukla, Time Refraction and Time Reflection: Two Basic Concepts. *Phys. Scr*. **65**, 160 (2002).

7. G. Lerosey, J. de Rosny, A. Tourin, A. Derode, G. Montaldo, M. Fink, Time reversal of electromagnetic waves. *Phys. Rev. Lett*. **92**.193904 (2004).

8. V. Bacot, M. Labousse, A. Eddi, M. Fink, E. Fort, Time reversal and holography with spacetime transformations. *Nat. Phys*. **12**, 972–977 (2016).

9. M. Fink, Time reversal of ultrasonic fields. I. Basic principles. *IEEE Trans. Ultrason Ferroelectr Freq. Control.* **39**, 555–66 (1992).

10. C. Draeger, M. Fink, One-channel time reversal of elastic waves in a chaotic 2D-silicon cavity. *Phys. Rev. Lett*. **79**, 407–410 (1997).

11. F. Biancalana, A. Amann, A. V. Uskov, E. P. O'Reilly, Dynamics of light propagation in spatiotemporal dielectric structures. *Phys. Rev. E* **75**, 046607.(2007)

12. J. R. Reyes-Ayona, P. Halevi, Observation of genuine wave vector ( k or β ) gap in a dynamic transmission line and temporal photonic crystals. *Appl. Phys. Lett*. 107, 074101 (2015).

13. A. B. Shvartsburg, Optics of nonstationary media, *Physics-Uspekhi* **48**, 797 (2005).

14. M. Uhlmann, G. Plunien, R. Schützhold, G. Soff, Resonant cavity photon creation via the dynamical Casimir effect. *Phys. Rev. Lett*. **93**.19 (2004).

15. J. T. Mendonca, G. Brodin, M. Marklund, Vacuum effects in a vibrating cavity: time refraction, dynamical Casimir effect, and effective Unruh acceleration. *Phys. Lett. A* **372**.35 (2008).

16. J. Sloan, N. Rivera, J. D. Joannopoulos, M. Soljačić, Casimir Light in Dispersive Nanophotonics. *Phys. Rev. Lett*. **127**, 053603.(2021)

17. F. Belgiorno, S. L. Cacciatori, M. Clerici, V. Gorini, G. Ortenzi, L. Rizzi, E. Rubino, V. G. Sala, and D. Faccio, Hawking radiation from ultrashort laser pulse filaments." *Phys. Rev. Lett.*, **105** (20), 203901 (2010).

18. E. Lustig, Y. Sharabi, M. Segev, Topological aspects of photonic time crystals. *Optica* **5**, 1390-1395 (2018).

19. V. Pacheco-Peña, N. Engheta, Antireflection temporal coatings. *Optica* **7**, 323-331 (2020).

20. H. Li, S. Yin, E. Galiffi, A. Alù, Temporal parity-time symmetry for extreme energy transformations. *Phys. Rev. Lett*. **127**(15). 153903.(2021)

21. Y. Sharabi, E. Lustig, M. Segev, Disordered Photonic Time Crystals, *Phys. Rev. Lett.*, **126** (16), 163902 (2021).

22. N. Kinsey, C. DeVault, J. Kim, M. Ferrera, V. M. Shalaev, A. Boltasseva, Epsilon-near-zero Al-doped ZnO for ultrafast switching at telecom wavelengths. *Optica* **2** 616-622 (2015).

23. L. Caspani, R. P. M. Kaipurath, M. Clerici, M. Ferrera, T. Roger, J. Kim, N. Kinsey, M. Pietrzyk, A. Di Falco, V. M. Shalaev, A. Boltasseva, D. Faccio, Enhanced nonlinear refractive index in ε-near-zero materials. *Phys. Rev. Lett.,* **116** (23), 233901 (2016).

24. M. Z. Alam, I. De Leon, R. W. Boyd. Large optical nonlinearity of indium tin oxide in its epsilon-near-zero region. *Science*, **352**.6287 795-797 (2016).

25. O. Reshef, I. De Leon, M. Z. Alam, R. W. Boyd, Nonlinear optical effects in epsilon-

near-zero media. *Nat Rev Mater* **4**, 535–551 (2019).

26. E. Lustig, S. Saha, E. Bordo, C. DeVault, S. N. Chowdhury, Y. Sharabi, A. Boltasseva, O. Cohen, V. M. Shalaev, M. Segev, Towards photonic time-crystals: observation of a femtosecond time-boundary in the refractive index. *2021 Conference on Lasers and Electro-Optics (CLEO)* (2021).

27. E. Yablonovitch, Inhibited spontaneous emission in solid-state physics and electronics. *Phys. Rev. Lett.* **58**, 2059–2062 (1987).

28. O. Painter, R. K. Lee, A. Scherer, A. Yariv, J. D. O'Brien, P. D. Dapkus, I. Kim, Two-dimensional photonic band-gap defect mode laser. *Science* **284**, 1819–1821 (1999).

29. S. Noda, C. Alongkarn, I. Masahiro, Trapping and emission of photons by a single defect in a photonic bandgap structure. *Nature* **407**, 608-610 (2000).

30. See supplementary material

31. G. T. Moore, Quantum theory of the electromagnetic field in a variable-length one-dimensional cavity. *J. Mat. Phys*. **11**, 2679 (1970).

32. The state of art in DCE is described in [33]. In the context of DCE, most works either utilize the Heisenberg representation of the system dynamics [15,34] and considers only small variations of refractive index so that photonic bandgap narrows down to one mode [37, 38], and/or considers the field in a cavity, imposing boundary conditions and automatically leaving only one mode from pair {k,-k } [15,35,36]

33. V. V. Dodonov, Current status of the dynamical Casimir effect. *Physica Scripta* **82**.3 038105 (2010).

34. C. Krattenthaler, S. I. Kryuchkov, A. Mahalov, S. Suslov, On the problem of electromagnetic-field quantization. *International Journal of Theoretical Physics* **52**.12 4445-4460 (2013).

35. C. K. Law, Effective Hamiltonian for the radiation in a cavity with a moving mirror and a time-varying dielectric medium. *Physical Review A* **49**.1 433 (1994).

36. C. K. Law, Interaction between a moving mirror and radiation pressure: A Hamiltonian formulation. *Physical Review A* **51**.3 (1995).

37. W. Vogel, D. -G. Welsch, Quantum optics (Wiley 2006).

38. X. -M. Bei, and Z. -Z. Liu. Quantum radiation in time-dependent dielectric media. *Jour. Phys. B:* Atomic, *Molecular and Optical Physics*, **44**.20, 205501 (2011).